\documentclass[12pt,showkeys,amsmath,amssymb]{revtex4}
\usepackage{amsmath,amsfonts,amsthm,amscd,amssymb,latexsym}
\usepackage{bm}
\usepackage{dcolumn}
\usepackage{graphicx}
\usepackage{epstopdf}
\usepackage{color}
\usepackage{epsf}
\usepackage{epsfig}
\usepackage{graphicx, epic, eepic, color}
\usepackage[colorlinks,citecolor=blue,urlcolor=blue,linkcolor=blue]{hyperref}

\begin{document}

\title{Schwarzschild Spacetime and the Local Limit of Nonlocal Gravity}

\author{Yahya \surname{Mohammadi}$^{1}$}
\email{yahya.mohammadi848@sharif.edu}
\author{Javad \surname{Tabatabaei}$^{1}$}
\email{smj\_tabatabaei@physics.sharif.edu}
\author{Shant \surname{Baghram}$^{1,2}$}
\email{baghram@sharif.edu}
\author{Bahram \surname{Mashhoon}$^{1,3,4}$} 
\email{mashhoonb@missouri.edu}

\affiliation{
$^1$Department of Physics, Sharif University of Technology, Tehran 11155-9161, Iran\\
$^2$Research Center for High Energy Physics,
Department of Physics,
Sharif University of Technology,
Tehran 11155-9161, Iran\\
$^3$School of Astronomy,
Institute for Research in Fundamental Sciences (IPM),
Tehran 19395-5531, Iran\\
$^4$Department of Physics and Astronomy,
University of Missouri, Columbia,
Missouri 65211, USA
}

\date{\today}

\begin{abstract}
We investigate static spherically symmetric solutions within the framework of the local limit of nonlocal gravity, which is a modified form of the teleparallel equivalent of general relativity. This theory departs from Einstein's general relativity through the introduction of a scalar gravitational susceptibility function $S(x)$, $1+S > 0$, which is a new phenomenological feature of the spacetime that vanishes in the general relativity limit.  It is shown that the Schwarzschild spacetime constitutes an exact solution of the modified source-free gravitational field equations provided $S$ is a constant.  The mass of the corresponding Schwarzschild solution is given by $M/(1+S)$, where $M$ denotes the mass of the solution in the general relativity limit. The interpretation of the solution in terms of a black hole is ruled out due to the divergence of the algebraic invariants of the Weitzenb\"ock torsion at the Schwarzschild horizon. 
\end{abstract}


\maketitle


\section{Introduction}

Nonlocal gravity (NLG) is a classical nonlocal generalization of general relativity (GR) modeled after the only successful classical nonlocal field theory, namely,  the nonlocal electrodynamics of media. This is a classical nonlocal generalization of Maxwell's original equations in which one retains Maxwell's equations for $(\mathbf{E}, \mathbf{B})$ and $(\mathbf{D}, \mathbf{H})$, but the constitutive relations between these fields are nonlocal. The underlying medium is an atomic system. To model NLG after the electrodynamics of nonlocal media, one must first express GR in a form that resembles Maxwell's electrodynamics. Indeed, there is a well-known teleparallel equivalent of general relativity (TEGR), which is the gauge theory of the Abelian group of spacetime translations. Therefore, TEGR, though nonlinear, is formally analogous to electrodynamics and can be rendered nonlocal via history-dependent constitutive relations as in the electrodynamics of nonlocal media. In the resulting theory of nonlocal gravity (NLG), the gravitational field is locally defined but satisfies partial integro-differential field equations. NLG employs an extended geometric framework involving a preferred orthonormal frame field; that is, it is a tetrad theory. The preferred tetrad field defines the metric. The spacetime manifold has two metric-compatible connections: the standard Levi-Civita connection of GR and the Weitzenb\"ock connection associated with the preferred frame field. The symmetric Levi-Civita connection is torsion free but has Riemannian curvature, while the Weitzenb\"ock connection is curvature free but has torsion. This extended framework is consistent with the universality of the gravitational interaction. Hence, free test particles and null rays follow geodesics of the geometry. Thus far, the nonlinearity of NLG has prevented finding exact solutions for strong-field regimes such as those involving black holes or cosmological models; in fact, the only known exact solution of NLG is the trivial solution (i.e., Minkowski spacetime in the absence of gravity).  However, linearized NLG and its Newtonian limit have been extensively studied, since in this regime the nonlocal aspect of gravity has been shown to simulate dark matter. As a field theory of gravitation, GR was patterned after electrodynamics; in fact, it is a unique feature of NLG that it is patterned after the nonlocal electrodynamic of media. In this connection, the nonlocal kernel of NLG must be determined on the basis of observational data, since no underlying atomic system is available in the gravitational case. In practice, the nonlocal constitutive relations of nonlocal electrodynamics are often simplified and expressed in their local limits as $\mathbf{D} = \epsilon(x) \mathbf{E}$ and $\mathbf{B} = \mu (x) \mathbf{H}$, where the $(\epsilon, \mu)$ functions represent characteristic features of the medium. Similarly, one can introduce a scalar function $S(x)$ as the local limit of the nonlocal kernel of NLG. With $S(x)$, we then have a simpler theory that is the \emph{local limit of NLG} and in which we can explore black holes as well as cosmological models. The significance of this model lies in its ability to reveal some of the nontrivial effects that nonlocality may introduce into gravitational physics. It constitutes a self-consistent modification of GR involving a free function $S(x)$ thus leading to richer astrophysical and cosmological phenomenology. 

The purpose of this study is to determine the status of the black hole (BH) solutions of GR~\cite{HaEl,  Stephani:2003tm, GrPo} within the context of the local limit of NLG. In this paper, we consider the simple case of the static and spherically symmetric Schwarzschild spacetime and show that it is a regular solution of the modified theory in the exterior region since scalar torsion invariants diverge on the horizon; therefore, it is excluded as a BH. 

In GR, we imagine a point mass $M$ at the origin of spatial spherical polar coordinates. The resulting spacetime is given by the standard Schwarzschild metric,
\begin{equation}\label{I1}
ds^2 = - \left(1-  \frac{2\mu_0}{\rho}\right) c^2 dt^2 + \frac{d\rho^2}{1-  \frac{2\mu_0}{\rho}}  + \rho^2 ( d\theta^2 + \sin^2\theta\, d\phi^2)\,, 
\end{equation}
where $\mu_0 := GM/c^2$. The coordinate system is admissible in the exterior region, i.e., $\rho > 2\,\mu_0$; however, the solution can be analytically extended to the interior region where $\rho = 0$ is a curvature singularity. The Schwarzschild horizon at $\rho = 2\,\mu_0$ is a regular null hypersurface, which permits the interpretation of this static solution as a BH, namely, the end state of the complete gravitational collapse in GR.    The geometry of the Ricci-flat Schwarzschild spacetime is well known~\cite{HaEl,  Stephani:2003tm, GrPo}. The curvature of Schwarzschild spacetime is of type D in the Petrov classification. 

In NLG, we work in Cartesian coordinates; therefore, we need to express the Schwarzschild metric in isotropic coordinates. The isotropic radial coordinate $r$ is related to Schwarzschild radial coordinate $\rho$ by
\begin{equation}\label{I2}
 \rho = r\,\left(1+\frac{\mu_0}{2r}\right)^2\,.
\end{equation}
Then, the Schwarzschild metric in isotropic coordinates is
\begin{equation}\label{I3}
ds^2 = -\frac{1}{A_{\rm Sch}^2}\,c^2 dt^2 + \frac{1}{B_{\rm Sch}^2} (dx^2+dy^2+dz^2)\,, \quad A_{\rm Sch}(r) = \frac{1+ \frac{\mu_0}{2r}}{1- \frac{\mu_0}{2r}}\,, \quad B_{\rm Sch}(r) = \frac{1}{\left(1+\frac{\mu_0}{2r}\right)^{2}}\,,
\end{equation}
where $x = r \sin \theta \cos\phi$, $y = r \sin \theta \sin\phi$ and $z = r \cos \theta$. The coordinate system $x^\mu = (ct, x, y, z)$ is admissible for the exterior Schwarzschild spacetime ($r > \mu_0/2$).  

From now on, we use natural units such that the speed of light in vacuum $c$ and Newton's constant of gravitation 
$G$ are set equal to unity, unless specified otherwise; moreover,  the metric signature is +2 and Greek indices run from 0 to 3, while Latin indices run from 1 to 3. 

Einstein's gravitational field equations are given by~\cite{Einstein}
\begin{equation}\label{I4} 
 {^0}G_{\mu \nu} + \Lambda g_{\mu \nu} = \kappa\,T_{\mu \nu}\,, \qquad \kappa:= \frac{8 \pi G}{c^4}\,,
\end{equation}
where ${^0}G_{\mu \nu}=~ ^0R_{\mu \nu} - \frac{1}{2} g_{\mu \nu}\, ^0R$ is the Einstein tensor.  Throughout this paper,  we employ a left superscript ``0" to refer to geometric quantities that are directly related to the Levi-Civita connection ($^0\Gamma^\mu_{\alpha \beta}$).  In Eq.~\eqref{I4}, $\Lambda$ is the cosmological constant and $T_{\mu \nu}$ is the energy-momentum tensor of the gravitational source. 

Henceforward, we consider a general static and spherically symmetric metric of the form
\begin{equation}\label{I5}
ds^2 = -\frac{1}{A^2(r)}\, dt^2 + \frac{1}{B^2(r)} (dx^2+dy^2+dz^2)\,. 
\end{equation}
It is straightforward to compute the Einstein tensor in this case and its nonzero components are given by
\begin{equation}\label{I6} 
 {^0}G_{00} = \frac{1}{A^2(r)}\, \mathbb{T}\,,
\end{equation}
\begin{equation}\label{I7} 
 {^0}G_{ij} = \frac{x^i\,x^j}{r^2}\, (\mathbb{R} + \mathbb{S}) - \mathbb{S}\, \delta_{ij}\,.
\end{equation}
Here, $\mathbb{T}$, $\mathbb{R}$ and  $\mathbb{S}$ are given by
\begin{equation}\label{I8} 
\mathbb{T} = 2 B B'' - 3 B'^2 + \frac{4}{r} BB'\,,
\end{equation}
\begin{equation}\label{I9} 
\mathbb{R} = \left(\frac{B'}{B}\right)^2 + 2 \frac{A'}{A}\,\frac{B'}{B} -\frac{2}{r}\left(\frac{A'}{A} + \frac{B'}{B}\right)\,,
\end{equation}
\begin{equation}\label{I10} 
\mathbb{S} = \frac{A''}{A} + \frac{B''}{B} -2 \left(\frac{A'}{A}\right)^2 - \left(\frac{B'}{B}\right)^2  +\frac{1}{r}\left(\frac{A'}{A} + \frac{B'}{B}\right)\,,
\end{equation}
where $A' := dA/dr$, etc.  Note that  for $(A , B) = (A_{\rm Sch},  B_{\rm Sch})$, we have ${^0}G_{\mu \nu} = 0$; conversely, the Ricci-flat condition, namely, 
\begin{equation}\label{I11} 
\mathbb{T} = 0\,, \qquad \mathbb{R} = 0\,, \qquad \mathbb{S} = 0\,,
\end{equation}
uniquely specifies the Schwarzschild solution. 

Does metric~\eqref{I5} fit into the framework of the local limit of NLG? We note that NLG is a classical nonlocal generalization of  the teleparallel equivalent of GR (TEGR)~\cite{Hehl:2008eu, Hehl:2009es}. Indeed, it is possible to express GR within the context of teleparallelism~\cite{We, BlHe, Maluf:2013gaa, Aldrovandi:2013wha, Itin:2018dru,BeltranJimenez:2019esp,Capozziello:2022zzh}.  The resulting teleparallel equivalent of GR, namely, TEGR, is the gauge theory of the Abelian group  of spacetime translations~\cite{Cho}. There is a formal analogy between the gravitational field equations of TEGR and Maxwell's equations for the electrodynamics of media. It is then possible to render TEGR nonlocal in a manner that is similar to the nonlocal electrodynamics of media~\cite{Hehl:2008eu, Hehl:2009es}.

To elucidate the connection between TEGR and NLG, let us note that in an inertial frame of reference, the field equations of Maxwell's electrodynamics of media include the electromagnetic field strength $(\mathbf{E}, \mathbf{B}) \mapsto F_{\mu\nu}$,
\begin{equation}\label{I12}
F_{\mu \nu} = \partial_{ \mu}A_{ \nu} -\partial_{ \nu}A_{ \mu}\,,
\end{equation}
where $A_\mu$ is the electromagnetic vector potential, and the electromagnetic field excitations $(\mathbf{D}, \mathbf{H}) \mapsto H_{\mu\nu}$. The latter fields involve the  polarizability and magnetizability of the medium in response to $F_{\mu \nu}$. Moreover,  
\begin{equation}\label{I13}
\partial_{\nu}H^{\mu \nu} = \frac{4 \pi}{c} J^{\mu}\,,
\end{equation}
where $J^\mu$ refers to the current of free electric charges and is a conserved 4-vector. Furthermore, there exist constitutive relations that connect $F_{\mu \nu}$  and $H_{\mu\nu}$. These relations, which describe specific electromagnetic properties of the material  medium are in general nonlocal~\cite{Hop, Poi, Jack, L+L, HeOb, VanDenHoogen:2017nyy}. In their local limits, they can be expressed in familiar forms
\begin{equation}\label{I14}
\mathbf{D} = \epsilon(x)\, \mathbf{E}\,, \qquad  \mathbf{B} = \mu(x) \,\mathbf{H}\,
\end{equation}
where $\epsilon$ and $\mu$ are the electric permittivity and magnetic permeability of the medium, respectively.

Let us digress here and note that there are indeed various other frameworks for nonlocal gravitation. Modified gravities in the form of nonlocal $f(R)$ theories have been studied with the aim of providing a consistent theoretical foundation  for comparison with general cosmological observations; see, for instance,~\cite{Nojiri:2026hij,Nojiri:2019dio,Nojiri:2017ncd} and the references cited therein. On the other hand, specific nonlocal cosmological models have been developed to explain cosmic acceleration without dark energy as well as structure formation without actual dark matter; see, for instance,~\cite{Deffayet:2025lwl,Deser:2019lmm,Tan:2018bfp,  Woodard:2018gfj, Maggiore:2014sia, Balakin:2022gjw,  Koshelev:2022bvg, Koshelev:2022olc, Jusufi:2023ayv, Bajardi:2024kea, Capozziello:2024qol} and the references therein. Within the class of nonlocal theories of gravitation, our adopted framework~\cite{Hehl:2008eu, Hehl:2009es}, modeled after the electrodynamics of nonlocal media, provides a unique classical method of constructing nonlocal gauge theories of gravitation. Nonlocal TEGR has been treated in detail and has significant implications for dark matter and dynamic dark energy; moreover, our approach can be extended to nonlocal Poincar\'e gauge theory of gravity~\cite{Blome:2010xn}. 

A brief outline of the derivation of TEGR, NLG and its local limit is presented in Sec. II.  In Secs. III and IV, we determine the field equations of the local limit of nonlocal gravity for metric~\eqref{I5}. We solve these equations in Sec. V. The resulting solution is eliminated as a BH in Sec. VI.  We conclude with a discussion of our results in Sec. VII.

\section{Local Limit of Nonlocal Gravity}

To describe the local limit of NLG, we begin with the tetrad formulation of GR~\cite{BlHe, Maluf:2013gaa}. Consider a smooth family of observers in spacetime  that carry an orthonormal tetrad frame field $\lambda^\mu{}_{\hat \alpha}(x)$. Here, $x^\alpha = (t,x^i)$, $i = 1, 2, 3$, is an admissible system of coordinates that cover the spacetime under consideration here. Moreover, $\lambda^\mu{}_{\hat 0}(x)$ is a unit timelike direction that is the four-velocity vector of an observer in the congruence, while the triad $\lambda^\mu{}_{\hat i}(x)$, $i = 1, 2, 3$, is the observer's local spatial frame. The spacetime metric in the tetrad framework is recovered from the orthonormality condition of the tetrad field,
\begin{equation}\label{M1}
g_{\mu \nu}(x) = \lambda_\mu{}^{\hat \alpha}(x)\, \lambda_\nu{}^{\hat \beta}(x)\, \eta_{\hat \alpha \hat \beta}\,.
\end{equation}
We use the tetrad field in order to define a new connection. Indeed, 
\begin{equation}\label{M2}
\Gamma^\mu_{\alpha \beta} :=\lambda^\mu{}_{\hat \rho }~\partial_\alpha\,\lambda_\beta{}^{\hat \rho}\,,
\end{equation}
defines the Weitzenb\"ock connection~\cite{We} upon which the scheme of teleparallelism is based. Associated with the Weitzenb\"ock connection is the covariant derivative operator $\nabla$ which satisfies $\nabla_\nu\,\lambda_\mu{}^{\hat \alpha}=0$. The frame field is covariantly constant; furthermore, the Weitzenb\"ock connection is curvature free. This means that the spacetime is a parallelizable manifold and the tetrad frames are everywhere parallel. This is the framework of teleparallelism; that is, two faraway vectors are thus considered parallel if their local tetrad components are the same. 
By establishing a globally flat, parallelizable manifold, this connection ensures that the tetrad fields act strictly as the gauge potentials of spacetime translations. Consequently, the torsion tensor generated by the Weitzenb\"ock  connection becomes the sole carrier of the gravitational field dynamics in TEGR.
The Weitzenb\"ock connection is metric-compatible, namely, $\nabla_\mu\,g_{\alpha \beta} = 0$, which follows from applying $\nabla$ to Eq.~\eqref{M1}.

In our extended GR scheme, we have two metric-compatible connections, namely, the Weitzenb\"ock connection $\Gamma^\mu_{\alpha \beta}$ and the symmetric Levi-Civia connection $^0\Gamma^\mu_{\alpha \beta}$, 
\begin{equation}\label{M3}
^0\Gamma^\mu_{\alpha \beta} = \frac{1}{2} g^{\mu \nu}\,(g_{\nu \alpha, \beta} + g_{\nu \beta, \alpha} - g_{\alpha \beta, \nu})\,.
\end{equation}
\textcolor{red}{As in GR,} free test particles and null rays follow the geodesic equation, 
\begin{equation}\label{M4}
\frac{d^2 x^\mu}{d\theta^2} +\, ^0\Gamma^\mu_{\alpha \beta}\, \frac{dx^\alpha}{d\theta}\,\frac{dx^\beta}{d\theta} = 0\,,
\end{equation}
where $\theta$ is the proper time (affine parameter) for a timelike (null) geodesic. 

To proceed further, we define the \emph{torsion} tensor of the Weitzenb\"ock connection, namely, 
\begin{equation}\label{M5}
C_{\mu \nu}{}^{\rho}=\Gamma^{\rho}_{\mu \nu}-\Gamma^{\rho}_{\nu \mu}=\lambda^\rho{}_{\hat \beta }\left(\partial_{\mu}\lambda_{\nu}{}^{\hat \beta }-\partial_{\nu}\lambda_{\mu}{}^{\hat \beta}\right)\,,
\end{equation}
and the corresponding \emph{contorsion} tensor 
\begin{equation}\label{M6}
K_{\mu \nu}{}^{\rho} =\, ^{0}\Gamma^{\rho}_{\mu \nu} - \Gamma^{\rho}_{\mu \nu}\,.
\end{equation}
These definitions result in proper tensor fields, since the difference between two connections on the same manifold is always a tensor. The torsion tensor $C_{\alpha \beta \gamma}$ is antisymmetric in its first two indices, $C_{\alpha \beta \gamma} = - C_{\beta \alpha \gamma}$, while the contorsion tensor $K_{\alpha \beta \gamma}$ is antisymmetric in its last two indices, $K_{\alpha \beta \gamma} = - K_{\alpha \gamma \beta}$. On the other hand, the new tensors are related through $\nabla_\mu\,g_{\alpha \beta} = 0$, which means 
\begin{equation}\label{M7}
g_{\alpha \beta, \gamma} =  \Gamma^\mu_{\gamma \alpha}\,g_{\mu \beta} + \Gamma^\mu_{\gamma \beta}\,g_{\mu \alpha}\,.
\end{equation}
One can then show via Eq.~\eqref{M3} and definitions Eqs.~\eqref{M5}--\eqref{M6} that 
\begin{equation}\label{M8} 
K_{\mu \nu \rho} = \frac{1}{2}\, (C_{\mu \rho \nu}+C_{\nu \rho \mu}-C_{\mu \nu \rho})\,.
\end{equation}

\subsection{General Relativity and Teleparallelism}

The Levi-Civita connection is the sum of the Weitzenb\"ock connection and its contorsion tensor in accordance with Eq.~\eqref{M6}. Therefore, the Riemann tensor, $^0R_{\alpha \beta \gamma \delta}$, may be rewritten entirely in terms of the quantities associated with the torsion tensor. To this end, we define the torsion vector $C_\mu$,
\begin{equation}\label{M9}
C_\mu := C^{\alpha}{}_{\mu \alpha} = - C_{\mu}{}^{\alpha}{}_{\alpha}\,,
\end{equation}
and the auxiliary torsion tensor $\mathfrak{C}_{\alpha \beta \gamma}$, 
\begin{equation}\label{M10}
\mathfrak{C}_{\alpha \beta \gamma} :=C_\alpha\, g_{\beta \gamma} - C_\beta \,g_{\alpha \gamma}+K_{\gamma \alpha \beta} = C_\alpha\, g_{\beta \gamma} - C_\beta \,g_{\alpha \gamma} +\frac{1}{2}\, (C_{\gamma \beta \alpha}+C_{\alpha \beta \gamma}-
C_{\gamma \alpha \beta})\,.
\end{equation}
It then turns out that the Einstein tensor can be expressed as 
\begin{align}\label{M11}
 {^0}G_{\mu \nu}=\frac{\kappa}{\sqrt{-g}}\Big[\lambda_\mu{}^{\hat{\gamma}}\,g_{\nu \alpha}\, \frac{\partial}{\partial x^\beta}\,\mathfrak{H}^{\alpha \beta}{}_{\hat{\gamma}}
-\Big(C_{\mu}{}^{\rho \sigma}\,\mathfrak{H}_{\nu \rho \sigma}
-\frac{1}{4}\,g_{\mu \nu}\,C^{\alpha \beta \gamma}\,\mathfrak{H}_{\alpha \beta \gamma}\Big) \Big]\,,
\end{align}
where the auxiliary torsion field $\mathfrak{H}_{\mu \nu \rho}$ is given by
\begin{equation}\label{M12}
\mathfrak{H}_{\mu \nu \rho}:= \frac{\sqrt{-g}}{\kappa}\,\mathfrak{C}_{\mu \nu \rho}\,.
\end{equation}

Using Eq.~\eqref{M11} in the gravitational field Eq.~\eqref{I4}, we obtain the TEGR field equation
\begin{equation}\label{M13}
\frac{\partial}{\partial x^\nu}\,\mathfrak{H}^{\mu \nu}{}_{\hat{\alpha}}+\frac{\sqrt{-g}}{\kappa}\,\Lambda\,\lambda^\mu{}_{\hat{\alpha}} =\sqrt{-g}\,(T_{\hat{\alpha}}{}^\mu + \mathfrak{T}_{\hat{\alpha}}{}^\mu)\,.
\end{equation}
Here, $\mathfrak{T}_{\mu \nu}$ is the traceless energy-momentum tensor of the gravitational field given by
\begin{equation}\label{M14}
\mathfrak{T}_{\mu \nu} := (\sqrt{-g})^{-1}\, (C_{\mu \rho \sigma}\, \mathfrak{H}_{\nu}{}^{\rho \sigma}-\tfrac{1}{4}  g_{\mu \nu}\,C_{\rho \sigma \delta}\,\mathfrak{H}^{\rho \sigma \delta})\,.
\end{equation}
The antisymmetry of $\mathfrak{H}^{\mu \nu}{}_{\hat{\alpha}}$ in its first two indices results in
\begin{equation}\label{M15}
\frac{\partial}{\partial x^\mu}\,\left[\sqrt{-g}\,(T_{\hat \alpha}{}^\mu - \tfrac{1}{\kappa}\,\Lambda\,\lambda^\mu{}_{\hat \alpha}+ \mathfrak{T}_{\hat \alpha}{}^\mu)\right] = 0\,.
\end{equation}
This is the law of conservation of total energy-momentum tensor in TEGR. The problem of gravitational energy in general relativity seems solvable within the TEGR framework~\cite{Maluf:2013gaa, Aldrovandi:2013wha, BlHe}. 

The path we have traced from GR to TEGR holds for any smooth tetrad field $\lambda^\mu{}_{\hat \alpha}$. All tetrad fields that correspond to the same spacetime metric are related by local Lorentz transformations and GR treats them as physically equivalent. As a result, the tetrad description carries a six-fold redundancy: three boosts and three rotations of the tetrad at each spacetime event leave the physics intact. A thorough discussion can be found in~\cite{BMB} and the references cited therein.

The forthcoming extension to \textcolor{red}{NLG} as well as its local limit is motivated by an analogy between TEGR and Maxwell's electrodynamics of media; therefore, it is useful to mention this similarity here.  Let us write the torsion tensor in the form
 $C_{\mu \nu}{}^{\hat \alpha} = \lambda_\rho{}^{\hat \alpha}\,C_{\mu \nu}{}^\rho$, which, using Eq.~\eqref{M5}, can be expressed as 
\begin{equation}\label{M16}
C_{\mu \nu}{}^{\hat \alpha} = \partial_{\mu}\lambda_{\nu}{}^{\hat \alpha}-\partial_{\nu}\lambda_{\mu}{}^{\hat \alpha}\,.
\end{equation}
For each  $\hat{\alpha}$, Eq.~\eqref{M16} resembles the expression for the electromagnetic field tensor $F_{\mu \nu}$ in Eq.~\eqref{I12}. Furthermore, the resemblance extends to $\mathfrak{H}^{\mu \nu}{}_{\hat \alpha}$ and $H^{\mu \nu}$ in Eqs.~\eqref{M13} and~\eqref{I13}, respectively. Let us note that Eq.~\eqref{M12} connects $\mathfrak{H}_{\mu \nu \rho}$ with $C_{\mu \nu \rho}$; hence, Eq.~\eqref{M12} can be regarded as the local constitutive relation of TEGR.

In electrodynamics, $F_{\mu \nu}$ vanishes if the vector potential $A_{\mu}$ is a pure gauge; similarly, the torsion tensor in Eq.~\eqref{M16} vanishes if the  corresponding vector potential $\lambda_{\mu}{}^{\hat \alpha}$ is a pure gauge, namely,  
$\lambda_{\mu}{}^{\hat \alpha}= \partial_\mu X^{\hat \alpha}$ for some functions $X^{\hat \alpha}$. In this case, the orthonormality relation~\eqref{M1} leads to the conclusion that the spacetime is flat, i.e., $^0R_{\alpha \beta \gamma \delta} = 0$. Conversely, a nonzero $^0R_{\alpha \beta \gamma \delta}$ indicates that the torsion tensor does not in general vanish. 
Thus, within this modified framework of general relativity, the Riemann curvature tensor associated with the symmetric Levi-Civita connection and the torsion tensor associated with the curvature-free Weitzenb\"ock connection constitute complementary features of the gravitational field~\cite{BMB}.

\subsection{NLG and its Local Limit}

In dealing with the electrodynamics of different media, the constitutive relation changes with the medium while Maxwell's equations of electromagnetism remain the same. The electromagnetic properties of a medium are encoded in the constitutive relation that is in general nonlocal. In contemplating a nonlocal extension of TEGR, we can draw inspiration from the nonlocal electrodynamics of media~\cite{Hehl:2008eu, Hehl:2009es}. That is, we can keep the field equation of TEGR, but adopt a nonlocal constitutive relation instead of Eq.~\eqref{M12}. To this end, $\mathfrak{H}_{\mu \nu \rho}$ on both sides of the TEGR field Eq.~\eqref{M13} is supplanted with $\mathcal{H}_{\mu \nu \rho}$ given by
\begin{equation}\label{M16a}
\mathcal{H}_{\mu \nu \rho} := \frac{\sqrt{-g}}{\kappa}(\mathfrak{C}_{\mu \nu \rho}+ N_{\mu \nu \rho})\,,
\end{equation}
where the nonlocal tensor $N_{\mu \nu \rho} = - N_{\nu \mu \rho}$ represents a specific spacetime average of the gravitational field in accordance with causality.
This tensor is directly inspired by the nonlocal part of the constitutive relation in electrodynamics~\cite{Jack, L+L}. In the gravitational context, it represents the nonlocal response of the background spacetime, appearing as an additional constitutive term that introduces the past history of the gravitational field into its present dynamics.
The presence of such nonlocality is expected to eliminate the 6-fold degeneracy inherent in TEGR and lead to a unique solution of the field equation of NLG involving the tetrad field $e^{\mu}{}_{\hat \alpha}$ that is adapted to a set of fundamental observers in spacetime. Therefore, henceforward,
\begin{equation}\label{M17}
\lambda^{\mu}{}_{\hat \alpha}|_{\rm TEGR} \to e^{\mu}{}_{\hat \alpha}|_{\rm NLG}\,,
\end{equation}
so that the basic field equation of NLG becomes
\begin{equation}\label{M18}
 \frac{\partial}{\partial x^\nu}\,\mathcal{H}^{\mu \nu}{}_{\hat{\alpha}}+\frac{\sqrt{-g}}{\kappa}\,\Lambda\,e^\mu{}_{\hat{\alpha}} =\sqrt{-g}\,(T_{\hat{\alpha}}{}^\mu + \mathcal{T}_{\hat{\alpha}}{}^\mu)\,,
\end{equation}
where $\mathcal{T}_{\mu \nu}$ is now the nonlocal traceless energy-momentum tensor of the gravitational field, namely, 
\begin{equation}\label{M18a}
\mathcal{T}_{\mu \nu} := (\sqrt{-g})^{-1}\, (C_{\mu \rho \sigma}\, \mathcal{H}_{\nu}{}^{\rho \sigma}-\tfrac{1}{4}  g_{\mu \nu}\,C_{\rho \sigma \delta}\,\mathcal{H}^{\rho \sigma \delta})\,.
\end{equation}
The law of conservation of the total energy-momentum, Eq.~\eqref{M15}, now takes the form
\begin{equation}\label{M19}
\frac{\partial}{\partial x^\mu}\,\Big[\sqrt{-g}\,(T_{\hat{\alpha}}{}^\mu + \mathcal{T}_{\hat{\alpha}}{}^\mu - \tfrac{1}{\kappa}\,\Lambda\,e^\mu{}_{\hat \alpha })\Big]=0\,.
 \end{equation}

It remains to determine how the nonlocality tensor $N_{\mu \nu \rho}$ relates to the gravitational field. In the framework of \textcolor{red}{NLG}, we suppose~\cite{Puetzfeld:2019wwo, Mashhoon:2022ynk} 
\begin{equation}\label{M20}
N_{\hat \mu \hat \nu \hat \rho}(x) = \int \mathcal{K}(x, x')\,\{\mathfrak{C}_{\hat \mu \hat \nu \hat \rho}(x')+ \check{p}\,[\check{C}_{\hat \mu}(x')\, \eta_{\hat \nu \hat \rho}-\check{C}_{\hat \nu}(x')\, \eta_{\hat \mu \hat \rho}]\} \sqrt{-g(x')}\, d^4x'\,.
\end{equation}
Here,  $\mathcal{K}(x, x')$ is the causal kernel of NLG and $\check{p}\ne 0$ is a constant dimensionless parameter. Moreover, $\check{C}_\mu$ is the torsion pseudovector, 
\begin{equation}\label{M21}
\check{C}_\mu :=\frac{1}{3!} C^{\alpha \beta \gamma}\,\eta_{\alpha \beta \gamma \mu}\,,
\end{equation}
where $\eta_{\alpha \beta \gamma \delta}$ denotes the Levi-Civita tensor.  The causal kernel incorporates the influence of the gravitational field's past history (``memory") into the present field equations, thereby introducing nonlocality into the theory. A detailed treatment of NLG can be found in~\cite{BMB}.  

It is interesting to determine how nonlocality modifies the GR field equation. The TEGR  gravitational field equation is equivalent to Einstein's field equation of GR. To determine the modified GR field equation, we can simply employ Eq.~\eqref{M16a} to replace $\mathfrak{H}_{\mu \nu \rho}$ in the Einstein tensor~\eqref{M11} with $\mathcal{H}_{\mu \nu \rho} -\tfrac{\sqrt{-g}}{\kappa}\, N_{\mu \nu \rho}$ and use Eq.~\eqref{M18} to get
\begin{equation}\label{M22}
^{0}G_{\mu \nu} + \Lambda g_{\mu \nu} = \kappa T_{\mu \nu}   +  Q_{\mu \nu} -  \mathcal{N}_{\mu \nu}\,,
\end{equation}
where $Q_{\mu \nu}$ is a traceless tensor given by
\begin{equation}\label{M23}
Q_{\mu \nu}:=\kappa\,\left(\mathcal{T}_{\mu \nu} - \mathfrak{T}_{\mu \nu}\right)=C_{\mu \rho \sigma} N_{\nu}{}^{\rho \sigma}-\frac 14\, g_{\mu \nu}\,C_{ \delta \rho \sigma}N^{\delta \rho \sigma}\,
\end{equation}
and we define a new tensor $\mathcal{N}_{\mu \nu}$, 
\begin{equation}\label{M24}
\mathcal{N}_{\mu \nu} := g_{\nu \alpha} e_\mu{}^{\hat{\gamma}} \frac{1}{\sqrt{-g}} \frac{\partial}{\partial x^\beta}\,(\sqrt{-g}N^{\alpha \beta}{}_{\hat{\gamma}})\,.
\end{equation} 

The modified GR field Eq.~\eqref{M22} contains 16 relations that can be employed to specify the 16 components of the tetrad frame field $e^\mu{}_{\hat \alpha}$. Ten of the 16 tetrad components determine the spacetime metric tensor $g_{\mu \nu}$, while the  other 6 are local Lorentz  degrees of freedom (i.e., boosts and rotations). In a similar way, the 10 symmetric components of the modified field Eq.~\eqref{M22} basically determine the spacetime metric of NLG, while the 6 antisymmetric components, $Q_{[\mu \nu]} = \mathcal{N}_{[\mu \nu]}$, are integral constraint equations. 

The field Eq.~\eqref{M22} of NLG involves highly nonlinear partial integro-differential relations that contain an unknown kernel $\mathcal{K}(x, x')$. Because of this complexity, it has not been possible to study the nonlinear regime of NLG in cases such as BHs or cosmological models. However, the linearized version of NLG and its Newtonian limit have been studied in detail; in this regime, it seems that the kernel can be ascertained using observational data as the nonlocal feature of gravity can effectively mimic dark matter~\cite{Rahvar:2014yta, Chicone:2015coa, Roshan:2021ljs, Roshan:2022zov, Roshan:2022ypk}. Indeed, it has been possible to show that in the Newtonian regime the implications of NLG are consistent with gravitational dynamics in spiral galaxies and clusters of galaxies~\cite{Rahvar:2014yta} as well as in the Solar System~\cite{Chicone:2015coa}; for a comprehensive account, see~\cite{BMB}.

 On the other hand, no exact nontrivial solutions of NLG have been found. Minkowski spacetime, with $e^{\mu}{}_{\hat \alpha} = \delta^\mu_\alpha$, remains the only exact solution known in the absence of gravity and any sources. Furthermore, it is known that de Sitter spacetime is not a solution of NLG~\cite{Mashhoon:2022ynk}.

It is worth noting that there are other models of nonlocal gravitation in which \textcolor{red}{BH} solutions have been studied~\cite{DAgostino:2025wgl, DAgostino:2025yej, Borah:2025tvw}. Nonlocal electrodynamics of media has provided the pattern for NLG, which is the only one of this kind among the available models. 

In the nonlocal electrodynamics of media, one frequently needs to simplify matters and use local relations $\mathbf{D} = \epsilon(x) \mathbf{E}$ and $\mathbf{B} = \mu (x) \mathbf{H}$ in place of the actual nonlocal constitutive relations. These local limits are presumed to capture key aspects of the more general nonlocal behavior. Similarly, in the context of NLG, we can consider its local limit as a way to investigate its nonlinear regime. Indeed, we assume that $\mathcal{K}(x, x')$ is proportional to the 4D Dirac delta function, namely, 
\begin{equation}\label{M25}
\mathcal{K}(x, x') := \frac{S(x)}{\sqrt{-g(x)}}\,\delta(x-x')\,,
\end{equation}
 where $S(x)$ is a scalar function that represents the gravitational \emph{susceptibility} function associated with the background spacetime~\cite{Tabatabaei:2023qxw, Tabatabaei:2023lec, Tabatabaei:2022tbq, Tabatabaei:2023iwc}.  Just as the electrical permittivity $\epsilon(x)$ and magnetic permeability $\mu(x)$ are constitutive properties of a medium in electrodynamics, $S(x)$ is a characteristic property of the gravitational ``medium".  
Physically, $S(x)$ acts as an effective feature that encapsulates the underlying microphysics of space and time. A primary objective of this phenomenological approach is to understand and constrain this susceptibility via astrophysical and cosmological observations.
 
In the local limit, Eq.~\eqref{M20} for the nonlocality tensor $N_{\mu \nu \rho}$ takes on the simple local form 
\begin{equation}\label{M26}
N_{\mu \nu \rho}(x) = S(x)\,[\mathfrak{C}_{\mu \nu \rho}(x) + \check{p}\,(\check{C}_\mu\, g_{\nu \rho}-\check{C}_\nu\, g_{\mu \rho})]\,
\end{equation}
and the local constitutive relation becomes
\begin{equation}\label{M27}
\mathcal{H}_{\mu \nu \rho} = \frac{\sqrt{-g}}{\kappa}[(1+S)\,\mathfrak{C}_{\mu \nu \rho}+ S\,\check{p}\,(\check{C}_\mu\, g_{\nu \rho}-\check{C}_\nu\, g_{\mu \rho})]\,.
\end{equation}
In this relation, we recover the TEGR and hence GR for $S = 0$, since Eq.~\eqref{M27} then reduces to Eq.~\eqref{M12}; moreover, for $S \ne 0$, we assume $1+S > 0$ in oder to maintain the physical connection between the local limit of NLG and GR.

In previous work on the local limit of NLG, the standard spatially homogeneous and isotropic cosmological models have been examined in connection with the $H_0$ tension~\cite{Tabatabaei:2022tbq, Tabatabaei:2023iwc}. Specifically, the Newtonian limit of this local modified theory ensures its consistency with Solar System tests through an appropriate redefinition of the effective Newtonian gravitational constant~\cite{Tabatabaei:2022tbq}. We have also extensively employed its linearized regime in the analysis of cosmological perturbations, providing the theoretical framework required to confront the model with observational data involving the cosmic microwave background (CMB) and baryon acoustic oscillations (BAO)~\cite{Tabatabaei:2023iwc}. Future investigations regarding gravitational-wave observations and weak-lensing measurements will heavily rely on these theoretical results. Moreover, the G\"odel universe turns out to be a solution of the local limit of NLG if $S$ is constant, which is in agreement with the spatial homogeneity of the G\"odel spacetime~\cite{MaMa}. The purpose of the present work is to investigate BH spacetimes; therefore, we employ metric~\eqref{I5} in the rest of this paper to compute  the 16 components of 
\begin{equation}\label{M28}
Q_{\mu \nu} -  \mathcal{N}_{\mu \nu} := \mathbb{M}_{\mu \nu}\,
\end{equation}
and utilize the result to solve field Eq.~\eqref{M22} in this case.  

\section{Metric~\eqref{I5} and Extended GR}

The man purpose of  Secs. III-V is to find the conditions under which metric~\eqref{I5} could also be a solution of the local limit of NLG.  

For the calculations regarding metric~\eqref{I5}, we collect here some useful relations:
\begin{equation}\label{T1} 
e^{\mu}{}_{\hat 0} = A(r) \delta^\mu_0\,, \qquad  e^{\mu}{}_{\hat i} = B(r) \delta^\mu_i\,,
\end{equation} 
\begin{equation}\label{T2} 
e_{\mu}{}^{\hat 0} = \frac{1}{A}\,\delta^0_\mu\,, \qquad  e_{\mu}{}^{\hat i} = \frac{1}{B}\,\delta^i_\mu\,.
 \end{equation}
Metric~\eqref{I5} in matrix form can be expressed as
\begin{equation}\label{T3}
(g_{\mu \nu}) = {\rm diag}(-A^{-2}, B^{-2}, B^{-2}, B^{-2})\,, \qquad (g^{\mu \nu}) = {\rm diag}(-A^2, B^2, B^2, B^2)\,.
\end{equation} 

In this case, the nonzero components of the Weitzenb\"ock connection are given by
\begin{equation}\label{T4} 
 \Gamma^0_{i0} = A\, \partial_i \frac{1}{A} = -\partial_i \ln A\,, \qquad   \Gamma^j_{ik} = B\, \partial_i \frac{1}{B} \delta^j_{k} = -\partial_i \ln B\,\delta^j_{k}\,.
\end{equation}
Let us note that $\partial_i F(r) = (F'/r) x^i$.  

The nonzero components of the torsion tensor are then given by
\begin{equation}\label{T5} 
 C_{0i}{}^0 = - C_{i0}{}^0 = \partial_i \ln A\,, \qquad   C_{jk}{}^i = -\partial_j \ln B\, \delta^i_k + \partial_k \ln B\, \delta^i_j\,. 
\end{equation}
Hence, the only nonzero components of $C_{\mu \nu \rho}$ are
\begin{equation}\label{T6} 
 C_{0i0} = -C_{i00} =  - \frac{1}{A^2} \,\partial_i \ln A\,, \qquad C_{jki} = \frac{1}{B^2}\,\left(-\partial_j \ln B\, \delta_{ik} + \partial_k \ln B\, \delta_{ij}\right)\,. 
\end{equation}
The torsion vector $C_{\mu} = C^{\alpha}{}_{\mu \alpha}$ has components
\begin{equation}\label{T7} 
 C_{0} = 0\,, \qquad C_i = \partial_i \ln (AB^2)\,, 
\end{equation}
while the torsion pseudovector vanishes, i.e. $\check{C}_{\mu} = 0$. 

The  nonzero components of the contorsion tensor~\eqref{M7} are given by
\begin{equation}\label{T8} 
 K_{0i0} = - K_{00i} = \frac{1}{A^2} \,\partial_i \ln A\,, \qquad K_{ijk} = \frac{1}{B^2}\,\left(\partial_k \ln B\, \delta_{ij} - \partial_j \ln B\, \delta_{ik}\right)\,. 
\end{equation}

The nonzero components of the auxiliary torsion tensor~\eqref{M10} can be expressed as 
\begin{equation}\label{T9} 
 \mathfrak{C}_{0i0} = - \mathfrak{C}_{i00} = \frac{2}{A^2} \,\partial_i \ln B\,, \qquad \mathfrak{C}_{jki} = \frac{1}{B^2}\,\left[\partial_j \ln (AB)\, \delta_{ik} - \partial_k \ln (AB)\, \delta_{ij}\right]\,,
\end{equation}
\begin{equation}\label{T10} 
 \mathfrak{C}^{0i0} = - \mathfrak{C}^{i00} = 2A^2\,B^2 \,\partial_i \ln B\,, \qquad \mathfrak{C}^{jki} = B^4\,\left[\partial_j \ln (AB)\, \delta_{ik} - \partial_k \ln (AB)\, \delta_{ij}\right]\,.
\end{equation}

We can now compute $Q_{\mu \nu}$, where
\begin{equation}\label{T11}
Q_{\mu \nu} := S(r)\,\left(C_{\mu \rho \sigma} \mathfrak{C}_{\nu}{}^{\rho \sigma}-\tfrac{1}{4}\, g_{\mu \nu}\,C_{ \delta \rho \sigma}\mathfrak{C}^{\delta \rho \sigma}\right)\,.
\end{equation} 
Let us note that
\begin{equation}\label{T12}
 C_{ \delta \rho \sigma}\mathfrak{C}^{\delta \rho \sigma} = -4 B^2\,\partial_i \ln B\,\partial_i \ln (A^2\,B)\,,
\end{equation} 
where, in our convention, repeated spatial indices, up or down, are summed over. After some algebra,  we find the nonzero components of the traceless tensor $Q_{\mu \nu}$, namely, 
\begin{equation}\label{T13}
{Q}_{00} = -S\,\frac{B^2}{A^2}\,\partial_i \ln B\,\partial_i \ln B\,,
\end{equation} 
\begin{equation}\label{T14}
{Q}_{ij} = -S\,[\partial_i \ln B\,\partial_j \ln (AB) + 2\, \partial_i \ln A\,\partial_j \ln B - \partial_k \ln A\,\partial_k \ln B\,\delta_{ij}]\,.
\end{equation} 

Next, we turn to the calculation of $\mathcal{N}_{\mu \nu}$,  
\begin{equation}\label{T15}
\mathcal{N}_{\mu \nu} := g_{\nu \alpha} e_\mu{}^{\hat{\gamma}} \frac{1}{\sqrt{-g}} \frac{\partial}{\partial x^\beta}\,(\sqrt{-g}N^{\alpha \beta}{}_{\hat{\gamma}})\,,
\end{equation} 
where $\sqrt{-g} = (AB^3)^{-1}$ and  
\begin{equation}\label{T16}
N^{\alpha \beta}{}_{\hat{\gamma}} = S(r) g^{\alpha \mu}\,g^{\beta \nu}\mathfrak{C}_{\mu \nu \rho}\, e^{\rho}{}_{\hat \gamma}\,.
\end{equation} 
Let us recall that $S(x)$ is a characteristic of the Schwarzschild spacetime whose metric only depends on the radial coordinate $r$; therefore, $S$ can only depend upon $r$. 

After some algebra, we finally find that the only nonzero elements of $\mathcal{N}_{\mu \nu}$ are given by
\begin{equation}\label{T17}
\mathcal{N}_{0 0} = 2\,\frac{B^3}{A^2}\,\partial_i \left(\frac{S}{B}\,\partial_i \ln B\right)\,,
\end{equation} 
\begin{equation}\label{T18}
\mathcal{N}_{ij} = A\,\partial_i \left[\frac{S}{A}\,\partial_j \ln (AB)\right] - A\,\partial_k \left[\frac{S}{A}\,\partial_k \ln (AB)\right]\,\delta_{ij}\,.
\end{equation}
 
We must now substitute these results in the modified gravitational field equation.

\section{Modified Gravitational Field Equation}

The field equation of the local limit of NLG for metric~\eqref{I5} is given by
\begin{equation}\label{F1}
^0G_{\mu \nu} + \Lambda g_{\mu \nu} = \kappa T_{\mu \nu}   +   \mathbb{M}_{\mu \nu}\,,
\end{equation} 
where $\mathbb{M}_{\mu \nu}$ is defined by Eq.~\eqref{M28}.

In the static situation under consideration here, we imagine the source is given by the energy-momentum tensor of a comoving perfect fluid, namely, 
\begin{equation}\label{F2a}
T_{\mu \nu} = (\varrho + p) u_\mu u_\nu + p\,g_{\mu \nu}\,,
\end{equation}
where $\varrho(r)$ is the energy density, $p(r)$ is the pressure and 
\begin{equation}\label{F2b}
 u^\mu = A(r) \,\delta^\mu_0 \,
\end{equation}
is the 4-velocity vector of the perfect fluid. Therefore, 
\begin{equation}\label{F2c}
(T_{\mu \nu}) = {\rm diag}\left(\varrho\,A^{-2}, p\,B^{-2},  p\,B^{-2}, p\,B^{-2}\right)\,.
\end{equation}
 
From the results of the previous section, we find the nonzero components of $\mathbb{M}_{\mu \nu}$, namely, 
\begin{equation}\label{F3}
\mathbb{M}_{00} = Q_{00} - \mathcal{N}_{00} = -\frac{1}{A^2} ( S\,\mathbb{T} + 2\, B B' S')\,,
\end{equation} 
\begin{equation}\label{F4}
\mathbb{M}_{ij} = Q_{ij} - \mathcal{N}_{ij} = \left(\delta_{ij} - \frac{x^i\,x^j}{r^2}\right)\,\left(\frac{A'}{A} + \frac{B'}{B}\right)\,S' - \left[\frac{x^i\,x^j}{r^2}\,(\mathbb{R} + \mathbb{S}) - \mathbb{S}\,\delta_{ij}\right] \,S\,,
\end{equation} 
where $\mathbb{T}$, $\mathbb{R}$ and  $\mathbb{S}$ are given by Eqs.~\eqref{I8}--\eqref{I10}. Therefore, employing Eqs.~\eqref{I6} and~\eqref{I7}, Eq.~\eqref{F1} reduces to 
\begin{equation}\label{F5}
\frac{1}{A^2}\,\mathbb{T} = \frac{\kappa \,\varrho + \Lambda}{A^2} -\frac{1}{A^2} ( S\,\mathbb{T} + 2\, B B' S')\,,
\end{equation} 
\begin{align}\label{F6}
\nonumber \frac{x^i\,x^j}{r^2}\,(\mathbb{R} + \mathbb{S}) - \mathbb{S}\,\delta_{ij} = {}&\frac{\kappa \,p - \Lambda}{B^2}\,\delta_{ij} + S'\left(\delta_{ij} - \frac{x^i\,x^j}{r^2}\right)\,\left(\frac{A'}{A} + \frac{B'}{B}\right)  \\
{}&- S\left[\frac{x^i\,x^j}{r^2}\,(\mathbb{R} + \mathbb{S}) - \mathbb{S}\,\delta_{ij}\right]\,.
\end{align} 
These equations are satisfied everywhere in space provided
\begin{equation}\label{F7}
(1+S)\,\mathbb{T} + 2\, B B' S' = \kappa \,\varrho + \Lambda\,,
\end{equation} 
\begin{equation}\label{F8}
 (1+S)\,(\mathbb{R} +\mathbb{S}) + \left(\frac{A'}{A} + \frac{B'}{B}\right)\,S' = 0\,,
\end{equation} 
\begin{equation}\label{F9}
 (1+S)\,\mathbb{S} + \left(\frac{A'}{A} + \frac{B'}{B}\right)\,S' = - \frac{\kappa \,p - \Lambda}{B^2}\,.
\end{equation} 


\section{Modified Schwarzschild Solution in Isotropic Coordinates} 

Let us now set $\varrho = p = \Lambda = 0$ and find the solution  of Eqs.~\eqref{F7}--\eqref{F9} in this case. We have a simple result using Eq.~\eqref{F7}; that is,  
\begin{equation}\label{F10}
(1+S)\,\left(2 B B'' - 3 B'^2 + \frac{4}{r} BB'\right)+ 2\, B B' S' = 0\,.
\end{equation} 
Next, subtracting Eq.~\eqref{F9} from Eq.~\eqref{F8} implies
\begin{equation}\label{F11} 
\mathbb{R} = \left(\frac{B'}{B}\right)^2 + 2 \frac{A'}{A}\,\frac{B'}{B} -\frac{2}{r}\left(\frac{A'}{A} + \frac{B'}{B}\right) = 0\,,
\end{equation}
which can be expressed as
\begin{equation}\label{F12} 
\left(\frac{A'}{A} + \frac{B'}{B}\right)^2 -\left(\frac{A'}{A}\right)^2  -\frac{2}{r}\left(\frac{A'}{A} + \frac{B'}{B}\right) = 0\,.
\end{equation}
Finally, using Eq.~\eqref{I10} for $\mathbb{S}$, we can write Eq.~\eqref{F9} in the form
\begin{equation}\label{F13} 
(1+S)\left[\left(\frac{A'}{A} + \frac{B'}{B}\right)'  - \left(\frac{A'}{A}\right)^2   +\frac{1}{r}\left(\frac{A'}{A} + \frac{B'}{B}\right)\right] + \left(\frac{A'}{A} + \frac{B'}{B}\right)\,S' = 0\,,
\end{equation}
which can be combined with Eq.~\eqref{F12} and written as
\begin{equation}\label{F14} 
\left(\frac{A'}{A} + \frac{B'}{B}\right)'  - \left(\frac{A'}{A} + \frac{B'}{B}\right)^2 +\left(\frac{A'}{A} + \frac{B'}{B}\right)\left( \frac{3}{r}+\frac{S'}{1+S}\right) = 0\,.
\end{equation}

Let us note that for $S = 0$,  field Eqs.~\eqref{F10},~\eqref{F12} and~\eqref{F14} are equivalent to Eq.~\eqref{I11}, which expresses the Ricci-flat condition of GR and uniquely results in the Schwarzschild solution. In the presence of the susceptibility function 
$S(r)$, we need to solve these field equations and determine the modified form of the Schwarzschild solution.  

It is reasonable to assume that  the spacetime under consideration must approach the Minkowski spacetime for $r \to \infty$; therefore, we assume $A(\infty) = B(\infty) = 1$. Dividing Eq.~\eqref{F10} by $BB'$, the resulting equation can be easily integrated and the solution is 
\begin{equation}\label{S1}
B = \frac{1}{(1+W)^2}\,, \qquad W(r):= - \frac{\mu }{2} \int_\infty^r \frac{du}{u^2[1+S(u)]}\,,
\end{equation} 
where $\mu$ is a constant. Next, from Eq.~\eqref{F11} we find
\begin{equation}\label{S2} 
\frac{A'}{A} = \frac{1}{2}\frac{B'}{B} \left(\frac{2B-rB'}{rB'-B}\right)\,,
\end{equation}
which, upon using Eq.~\eqref{S1}, becomes 
\begin{equation}\label{S3} 
\frac{A'}{A} = \left( \frac{2W'}{W+1}\right)\,\frac{rW' + W +1}{2rW' + W + 1}\,.
\end{equation}
Let us now consider 
\begin{equation}\label{S4} 
 \frac{B'}{B} =   -  \frac{2W'}{W+1}\,
\end{equation}
and define $\mathbb{U}(r)$, 
\begin{equation}\label{S5} 
\mathbb{U} := \frac{A'}{A} + \frac{B'}{B} =   - \frac{2rW'^2}{(W+1)(2rW' + W + 1)}\,.
\end{equation}
Note that Eq.~\eqref{F14} can be expressed as 
\begin{equation}\label{S6} 
\frac{d}{dr} \left(\frac{1}{\mathbb{U}}\right) -\frac{1}{ \mathbb{U}} \left( \frac{3}{r}+\frac{S'}{1+S}\right) + 1 = 0\,.
\end{equation}
Substituting $\mathbb{U}$ from  Eq.~\eqref{S5} in Eq.~\eqref{S6}, we find 
\begin{equation}\label{S7} 
\frac{d}{dr} \left(\frac{1}{\mathbb{U}}\right) -\frac{1}{ \mathbb{U}} \left( \frac{3}{r}+\frac{S'}{1+S}\right) + 1 = -\frac{2}{\mu^2} r^3 (W+1)^2 (S+1) S'\,,
\end{equation} 
which means that
\begin{equation}\label{S8} 
 \frac{dS}{dr} = 0\,. 
\end{equation}
Hence, we immediately find from Eqs.~\eqref{S1} and~\eqref{S3}, 
\begin{equation}\label{S9} 
A(r) = \frac{1+ \frac{\mu_S}{2r}}{1- \frac{\mu_S}{2r}}\,,\qquad B(r)  = \frac{1}{\left(1+\frac{\mu_S}{2r}\right)^{2}}\,, 
\end{equation} 
where
\begin{equation}\label{S10}
\mu_S = \frac{\mu}{1+S}\,.
\end{equation}
For $S = 0$, we must recover the standard GR result; hence,  $\mu = \mu_0 = G M/c^2$. 

The resulting metric is thus 
\begin{equation}\label{S11}
ds^2 = - \left(\frac{1- \frac{\mu_S}{2r}}{1+ \frac{\mu_S}{2r}}\right)^2\, dt^2 + \left(1+\frac{\mu_S}{2r}\right)^{4} (dx^2+dy^2+dz^2)\,. 
\end{equation}
Once we have the metric in the modified theory, we can change the isotropic coordinates back to the standard Schwarzschild coordinate system via 
\begin{equation}\label{S12}
 \rho_S = \left(1+\frac{\mu_S}{2r}\right)^2 \,r\,,
\end{equation}
so that 
\begin{equation}\label{S13}
ds^2 = - \left(1-  \frac{2\mu_S}{\rho_S}\right) c^2 dt^2 + \frac{d\rho_S^2}{1-  \frac{2\mu_S}{\rho_S}}  + \rho_S^2 ( d\theta^2 + \sin^2\theta\, d\phi^2)\,.
\end{equation}


\section{Weitzenb\"ock Invariants}

There are three independent algebraic torsion invariants given by~\cite{BMB}
\begin{equation}\label{W1}
 I_1 = C_{\alpha \beta \gamma}\,C^{\alpha \beta \gamma}\,, \qquad I_2 = C_{\alpha \beta \gamma}\,C^{\gamma \beta \alpha}\,,\qquad I_3 = C_{\alpha}\,C^{\alpha}\,.
\end{equation}
These scalars are built from quadratic contractions of the torsion tensor and appear in the Lagrangian of teleparallel theories.  The Weitzenb\"ock invariants represent facets of the strength of the gravitational field in TEGR.  They are similar to GR's Kretschmann scalar ($R_{\alpha\beta\gamma\delta}R^{\alpha\beta\gamma\delta}$), which provides a true coordinate-independent measure of the gravitational field.

The three invariants can be calculated for metric~\eqref{I5} via the nonzero components of the torsion tensor given in Sec. III. The results are
\begin{equation}\label{W2}
 I_1 = 2\,I_2 = 2 B^2 \left[ \left(\frac{A'}{A}\right)^2 +  \left(\frac{B'}{B}\right)^2\right] \,, \qquad  I_3 = B^2 \left(\frac{A'}{A} + 2\,\frac{B'}{B}\right)^2\,.
\end{equation}
Using the explicit expressions for $A(r)$ and $B(r)$ in Eq.~\eqref{S9}, we find
\begin{equation}\label{W3}
I_1 = 2\,I_2 = \mu_S^2\, \frac{r^2(6r^2-4\,\mu_S\, r + \mu_S^2)}{(r +\tfrac{\mu_S}{2})^6 (r -\tfrac{\mu_S}{2})^2}\,
\end{equation}
and 
\begin{equation}\label{W4}
I_3 = \mu_S^2\, \frac{r^2(r-\mu_S)^2}{(r +\tfrac{\mu_S}{2})^6 (r -\tfrac{\mu_S}{2})^2}\,.
\end{equation}

For $r > \tfrac{\mu_S}{2}$, these scalar invariants are positive and vanish as $r ^{-4}$ for $r \to \infty$; on the other hand, they all diverge as  $(r -\tfrac{\mu_S}{2})^{-2}$ on the horizon of the Schwarzschild spacetime. The horizon is thus singular  and the interior region $r < \tfrac{\mu_S}{2}$ is  not physically accessible. Therefore, only the exterior Schwarzschild spacetime for $r > \tfrac{\mu_S}{2}$ is physically significant.
 
It is important to emphasize that within the framework of teleparallel gravity, torsion invariants frequently diverge at black hole horizons. Specifically, in TEGR, it has been demonstrated that tetrad fields representing the Schwarzschild metric exhibit genuine torsion invariant divergences at the horizon surface, regardless of the chosen tetrad frame~\cite{Obukhov:2002tm, Obukhov:2019fti, Golovnev:2023uqb, Coley:2025cdj, Lopez:2025gvj}. Because our modified framework restricts the geometry to a preferred fundamental observer frame field, this divergence is a basic structural feature and cannot be dismissed as a tetrad-dependent artifact.

To illustrate simple applications of the local limit of NLG, two modified interior solutions of the exterior Schwarzschild solution of mass $M/(1+S)$ are considered in the appendices.


\section{DISCUSSION}

The Schwarzschild spacetime is a vacuum solution of the gravitational field equation of the local limit of NLG; however, the three Weitzenb\"ock torsion invariants diverge on the event horizon. This circumstance renders the event horizon a closed null hypersurface that is a naked singularity. Similar problems occur for BH candidates within the framework of teleparallel gravity theories~\cite{Obukhov:2002tm, Obukhov:2019fti, Golovnev:2023uqb, Coley:2025cdj, Lopez:2025gvj}. An analogous situation is encountered in GR when the Schwarzschild BH is slightly deformed by the addition of a small quadrupole moment. In this case, the event horizon becomes a naked singularity in agreement with the BH uniqueness theorems. A singular null hypersurface is still a one-way membrane and the requirement of causality can be satisfied in the exterior region. Indeed, quasinormal modes (QNMs) of such deformed collapsed configurations have been calculated in~\cite{ Allahyari:2018cmg, Allahyari:2019umx}. In principle, one can use observational data regarding QNMs to determine the additional quadrupole moment. In the local limit of NLG, the QNMs of the corresponding Schwarzschild configuration will depend on the constant gravitational susceptibility $S$ through the mass parameter $M/(1+S)$, which may have implications for the problem of the BH mass distribution~\cite{Kovetz:2016kpi}.  

For constant S, the exterior metric is exactly Schwarzschild with an effective mass $M/(1+S)$. Thus, classical tests that depend only on the exterior metric are degenerate with GR, differing merely by a rescaled mass. This degeneracy can be broken only by complementing astrophysical studies with a cosmological one that constrains $S$. Cosmological studies~\cite{Tabatabaei:2023lec, Tabatabaei:2022tbq} show that $S$ depends upon cosmic evolution. Therefore, although the present analysis assumed a static setting, a more complete treatment would likely introduce time dependence into the effective observed mass of compact objects over the universe's history. Such a feature could have significant implications for the phenomenology of these systems.

Several directions for future investigation emerge from the present work. First, it would be worthwhile to extend the analysis beyond the static regime and consider dynamical compact-object solutions, allowing the susceptibility field ($S$) to evolve in time~\cite{Tabatabaei:2024juy}. Such an effect could lead to novel observational signatures and place additional constraints on the theory. On the other hand, investigating the stability, QNM spectrum, and observational consequences of these solutions in astrophysical environments may provide further insight into the viability of this gravitational model as an extension of general relativity, in addition to the purely cosmological analysis we have pursued so far.

Finally, let us note that in contrast to the teleparallel gravity theories, BH solutions with regular torsion do exist in the Poincar\'e gauge theory (PGT)~\cite{Obukhov:2019fti}. Spacetime translations form a subgroup of the Poincar\'e group. Therefore, TEGR, the gauge theory of the translation group, is a subtype of PGT. It has been shown in Appendix C of Ref.~\cite{Blome:2010xn} that our unique approach to the nonlocal generalization of TEGR, patterned after the nonlocal electrodynamics of media,  can be extended to the PGT.  As in NLG, we do not expect to find exact nontrivial solutions of the nonlinear and nonlocal field equations of PGT.  We must therefore resort again to the local limit of nonlocal PGT and then search for BH solutions with regular torsion. This remains a task for the future.


 \section*{Acknowledgments}
 
BM is grateful to Friedrich Hehl and Yuri Obukhov for helpful discussions. We would like to thank Masoud Molaei for useful comments.
SB is partially supported by the Abdus Salam International Center for Theoretical Physics (ICTP) under the regular associateship scheme.
Moreover, JT and SB are partially supported by the Sharif University of Technology Office of Vice President for Research under Grant No. G4010204. 
 

\appendix

\section{Modified Interior Schwarzschild Solution}

It is interesting to match the exterior solution~\eqref{S13} with a certain modified interior Schwarzschild solution. Historically, the Interior Schwarzschild Solution~\cite{Stephani:2003tm, GrPo} is the first such solution with $\Lambda= 0$ related to 
metric~\eqref{I1}. This standard interior solution is given by
\begin{equation}\label{Z1}
ds^2 = -\frac{1}{4} \left( 3\,\sqrt{1-\frac{2\mu_0}{\rho_0}} - \sqrt{1-\frac{2\,\mu_0\, \rho^2}{\rho_0^3}}\right)^2 dt^2 + \frac{d\rho^2}{1-\frac{2\mu_0 \rho^2}{\rho_0^3}} + \rho^2 ( d\theta^2 + \sin^2\theta\, d\phi^2)\,, 
\end{equation}
which matches the exterior solution~\eqref{I1} continuously at the boundary surface $\rho = \rho_0$. The interior of this sphere has constant density
\begin{equation}\label{Z2}
\varrho = \frac{\mu_0}{\tfrac{4\pi}{3}\,\rho_0^3}\,,
\end{equation}
but the radial pressure monotonically increases from zero at the boundary to a maximum at the center, 
\begin{equation}\label{Z3}
p(\rho) = \varrho \, \frac{\cos \eta - \cos \eta_0}{3 \cos \eta_0 - \cos \eta}\,,
\end{equation}
provided
\begin{equation}\label{Z4}
\frac{2\mu_0}{\rho_0}  < \frac{8}{9}\,.
\end{equation}
Here, 
\begin{equation}\label{Z5}
\frac{\rho}{R_0} := \sin \eta\,, \qquad   \frac{\rho_0}{R_0} :=\sin\eta_0\,, \qquad R_0 := \left(\frac{\rho_0^3}{2\mu_0}\right)^{\frac{1}{2}}\,.
\end{equation}
The interior coordinate system is admissible for $\rho < R_0$. 

It turns out that the Interior Schwarzschild Solution is conformally flat~\cite{Stephani:2003tm, GrPo}.

Finally, it is straightforward to change $\mu_0$ to $\mu_S$ and $\rho$ to $\rho_S$ in the interior Schwarzschild solution to find an interior solution to metric~\eqref{S13}.  

\subsection{Modified Interior Schwarzschild Solution in Isotropic Coordinates} 

To illustrate Eqs.~\eqref{F7}--\eqref{F9} in the case of a perfect fluid, it is interesting to discuss the interior Schwarzschild solution in isotropic coordinates. We start with the standard GR solution~\eqref{Z1} and  introduce a new isotropic radial coordinate $r$ such that 
\begin{equation}\label{Z6}
r = \frac{\rho}{ 1 + \sqrt{1-\frac{\rho^2}{R_0^2}}}\,, \qquad \rho = \frac{2r}{1 + \frac{r^2}{R_0^2}}\,. 
\end{equation}
Then, in terms of the new radial coordinate we have
\begin{equation}\label{Z7}
ds^2 = -\frac{1}{4} \left( 3\cos \eta_0 - \cos \eta \right)^2 dt^2 + \frac{4}{\left(1+\frac{r^2}{R_0^2}\right)^2} (dx^2 + dy^2 + dz^2)\,, 
\end{equation}
where, as before,  $x = r \sin \theta \cos\phi$, $y = r \sin \theta \sin\phi$ and $z = r \cos \theta$. Moreover, 
\begin{equation}\label{Z8}
\cos \eta = \sqrt{1-\frac{\rho^2}{R_0^2}} = \frac{1-\frac{r^2}{R_0^2}}{1+\frac{r^2}{R_0^2}}\,, \qquad \cos \eta_0 = \sqrt{1-\frac{\rho_0^2}{R_0^2}} = \frac{1-\frac{r_0^2}{R_0^2}}{1+\frac{r_0^2}{R_0^2}}\,,
\end{equation}
\begin{equation}\label{Z9}
r_0 = \frac{\rho_0}{ 1 + \sqrt{1-\frac{\rho_0^2}{R_0^2}}}\,, \qquad \rho_0 = \frac{2r_0}{1 + \frac{r_0^2}{R_0^2}}\,, \qquad \frac{r_0^2}{R_0^2} < \frac{1}{2}\,
\end{equation}
and 
\begin{equation}\label{Z10}
\frac{8 \pi}{3} \, p(r) = \frac{r_0^2-r^2}{R_0^4 + 2 R_0^2 (r^2-r_0^2) -r^2r_0^2}\,.
\end{equation}
The pressure vanishes at $r = r_0$, the boundary of the perfect fluid sphere, where it should match to the exterior solution in isotropic coordinates. To bring this about, we let $t \to \mathcal{C} \,t$ and $(x, y, z) \to (\mathcal{D} x, \mathcal{D} y, \mathcal{D} z)$ in the \emph{exterior} coordinate system and determine $\mathcal{C}$ and $\mathcal{D}$ from the matching conditions based on the continuity of the metric at $r = r_0$.  

Finally, it is straightforward to change $\mu_0 \to \mu_0/ (1+S)$, where $S$ is a constant, in the interior  Schwarzschild solution in isotropic coordinates and show that Eqs.~\eqref{F7}--\eqref{F9} are indeed satisfied and the resulting modified interior solution matches the modified exterior solution. 


\section{Modified Oppenheimer-Snyder Model}

Let us now consider another situation that involves an interior Schwarzschild solution of the local limit of NLG. That is, we consider the standard Oppenheimer-Snyder model~\cite{Oppenheimer:1939ue, Datt:1938uwc, Khodabakhshi:2025fmf}. We show that the GR model goes over to the modified theory with a simple renormalization of mass, namely, $M \to M/(1+S)$ for constant $S$.  However, the contraction of the interior solution cannot proceed past the exterior Schwarzschild horizon since it is a singular null hypersurface.

\subsection{FLRW Models}

For the sake of completeness, we briefly describe in this section  the standard FLRW cosmological models given by the metric
\begin{equation}\label{H1}
 ds^2 = -dT^2 + \frac{a^2(T)}{K^2(r)}\,\delta_{ij}\,dx^i\,dx^j\,, 
\end{equation} 
where $a(T)$ is the scale factor, 
\begin{equation}\label{H2}
K(r)  = 1 + \frac{k}{4R_0^2 }\, r^2\,, \qquad  r^2 = \delta_{ij}\, x^i x^j\,, 
\end{equation}  
and $k = 1, -1$, or $0$,  for the closed, open, or flat model, respectively. Here, $R_0$ is a characteristic cosmological length scale. 

Let us define a new radial coordinate $R$,
\begin{equation}\label{H3}
R  := \frac{r}{1 + \frac{k}{4R_0^2 }\, r^2}\,;
\end{equation}  
then,  metric~\eqref{H1} takes the form
\begin{equation}\label{H4}
 ds^2 = -dT^2 + a^2(T)\,\left(\frac{dR^2}{1 - \frac{k\, R^2}{R_0^2 }} + R^2 d\Omega^2\right)\,. 
\end{equation} 

The source of the gravitational field is a perfect fluid of uniform density and pressure comoving with the preferred observers. The field equations in this case can be worked out using metric~\eqref{H1} and the nonzero components of the  Einstein tensor in the absence of the cosmological constant are given by 
\begin{equation}\label{H5}
G_{00} = \kappa \,\mu_{\rm FLRW} = 3\, \left(\frac{\dot{a}}{a}\right)^2 + 3\,\frac{k}{a^2\,R_0^2}\,,
\end{equation}
where $\dot{a} := da/dT$, and
\begin{equation}\label{H6}
G_{ij} = \kappa\, p_{\rm FLRW}\, \frac{a^2(t)}{K^2(r)}\,\delta_{ij} = -\frac{1}{K^2}(2a\,\ddot{a} + \dot{a}^2 + k/R_0^2)\,\delta_{ij}\,.
\end{equation}

\subsection{Closed Model}

Henceforth, we assume a closed model ($k = 1$) with density $\mu_{\rm FLRW}$ and zero pressure. The coordinate system is admissible for $R < R_0$; moreover,  these assumptions mean
\begin{equation}\label{H7}
\left(\frac{\dot{a}}{a}\right)^2 +\frac{1}{a^2\,R_0^2} = \frac{8 \pi}{3} \,\mu_{\rm FLRW}\,, \qquad \mu_{\rm FLRW} a^3 = C\,,
\end{equation}
where $C$ is a constant. Next, it is simple to show that in this case the Misner-Sharp mass function~\cite{Tabatabaei:2024juy} is given by
\begin{equation}\label{H8}
m(T, R) = \frac{4\pi}{3}(aR)^3\mu_{\rm FLRW} =  \frac{1}{2} \frac{R^3}{R_0^2}\,.
\end{equation}
This means that 
\begin{equation}\label{H9}
C = \frac{3}{8 \pi}\, \frac{1}{R_0^2}\,.
\end{equation}  
In this way, we find 
\begin{equation}\label{H10}
a(\eta) = \frac{1}{2}\, (1+\cos \eta)\,, \qquad  T(\eta) = \frac{1}{2}\, (\eta+\sin \eta) R_0\,.
\end{equation} 

\subsection{Interior Schwarzschild Solution} 

Note that $\dot{a} = 0$ at $a =1$ and $T= 0$. The universe starts to contract at this point and goes all the way to a singularity: $a = 0$ at $T = \pi \,R_0/2$. We take this solution to be the interior of an exterior Schwarzschild metric given by
\begin{equation}\label{H11}
ds^2 = - \left(1-  \frac{2\mathbb{M}}{r}\right) c^2 dt^2 + \frac{dr^2}{1-  \frac{2\mathbb{M}}{r}}  + r^2 ( d\theta^2 + \sin^2\theta\, d\phi^2)\,, 
\end{equation}
where $\mathbb{M}$ is the mass of Schwarzschild solution. Suppose that the boundary radius of the interior solution is at $R_b$, such that 
\begin{equation}\label{H12}
m(T, R_b) =   \frac{1}{2} \frac{R_b^3}{R_0^2}\,, \qquad R_0^2 = \frac{R_b^3}{2 m(T, R_b)}\,,
\end{equation}
where $m(T, R_b)$ is the total mass of the interior solution. We must now join the interior and exterior metrics. To this end, we introduce 
\begin{equation}\label{H13}
t = F(T, R)\,, \qquad r = G (T, R)\,,
\end{equation}
and rewrite metric~\eqref{H11} in the form
\begin{equation}\label{H14}
ds^2 = - \left(1-  \frac{2\mathbb{M}}{G}\right)(F_T dT + F_R dR)^2 + \frac{(G_T dT + G_R dR)^2}{1-  \frac{2\mathbb{M}}{G}}  + G^2 ( d\theta^2 + \sin^2\theta\, d\phi^2)\,. 
\end{equation}
Comparing this result with Eq.~\eqref{H4} for $k=1$, we conclude that we must have~\cite{MP} at $R = R_b$, 
\begin{equation}\label{H15}
 - \left(1-  \frac{2\mathbb{M}}{G}\right) F_T^2 + \frac{G_T^2}{1-  \frac{2\mathbb{M}}{G}} = -1\,, 
\end{equation}
\begin{equation}\label{H16}
 - \left(1-  \frac{2\mathbb{M}}{G}\right) F_R^2 + \frac{G_R^2}{1-  \frac{2\mathbb{M}}{G}} = \frac{a^2}{1 - \frac{R^2}{R_0^2 }}\,, 
\end{equation}
\begin{equation}\label{H17}
 \left(1-  \frac{2\mathbb{M}}{G}\right) F_T\,F_R = \frac{G_T\,G_R}{1-  \frac{2\mathbb{M}}{G}}\,, 
\end{equation}
\begin{equation}\label{H18}
G = a(T)\,R\,, \qquad G_R = a(T)\,. 
\end{equation}

From Eqs.~\eqref{H15} and~\eqref{H16}, we find
\begin{equation}\label{H19}
 \left(1-  \frac{2\mathbb{M}}{G}\right)^4 F_T^2\,F_R^2 = a^2(T)\, \left(1 - \frac{R_b^2}{R_0^2}\right)^{-1}\left(\dot{a}^2R_b^2 +1-  \frac{2\mathbb{M}}{aR_b}\right)\,\left(\frac{2\mathbb{M}}{aR_b}- \frac{R_b^2}{R_0^2}\right)\,. 
\end{equation}
On the other hand, Eq.~\eqref{H17} implies
\begin{equation}\label{H20}
 \left(1-  \frac{2\mathbb{M}}{G}\right)^4 F_T^2\,F_R^2 = \dot{a}^2 \,a^2(T)\, R_b^2\,. 
\end{equation}
From equating these results, we find
\begin{equation}\label{H21}
\dot{a}^2 + \frac{1}{R_0^2} = \frac{2\mathbb{M}}{a(T)R_b^3}\,. 
\end{equation}
Comparing this result with Eqs.~\eqref{H7} and~\eqref{H12}, we conclude 
\begin{equation}\label{H22}
 \frac{1}{R_0^2} = \frac{2\mathbb{M}}{R_b^3}\,, \qquad m(T, R_b) = \mathbb{M}\,. 
\end{equation}
That is, the exterior and interior masses must be the same. 

Finally, it is interesting to note that in~\cite{Tabatabaei:2022tbq}, it was shown that the closed FLRW model is a solution of the local limit of nonlocal gravity provided $S$ is a constant. Then, the density and pressure are simply modified by a factor of $1+S$, 
namely, $\varrho \to \varrho/(1+S)$ and $p \to p/(1+S)$. We then conclude that the Oppenheimer-Snyder model goes over to the local limit of NLG when we simply modify masses by dividing them by $1+S$. However, the solution is valid only to the extent that $r > 2 \mathbb{M}$, or $G = a(T) R > 2\,\mathbb{M}$.




\begin{thebibliography}{00}

\bibitem{HaEl}
S.~W. Hawking and G.~F.~R. Ellis, 
\emph{The Large Scale Structure of Space-Time} (Cambridge University Press, Cambridge, UK, 1973).

\bibitem{Stephani:2003tm}
H.~Stephani, D.~Kramer, M.~A.~H.~MacCallum, C.~Hoenselaers and E.~Herlt,
\emph{Exact Solutions of Einstein's Field Equations}, 2nd edn
(Cambridge University Press, Cambridge, UK, 2003).

\bibitem{GrPo}
J.~B. Griffiths and J. Podolsky,
\emph{Exact Space-Times in Einstein's General Relativity}
(Cambridge University Press, Cambridge, UK, 2009).


\bibitem{Einstein}
A. Einstein,  
\emph{The Meaning of Relativity} 
(Princeton University Press, Princeton, NJ, USA, 1955).


\bibitem{Hehl:2008eu}
F. W. Hehl and B. Mashhoon, 
 ``Nonlocal Gravity Simulates Dark Matter",
Phys. Lett. B \textbf{673}, 279-282 (2009).  
[arXiv:0812.1059 [gr-qc]]

          

\bibitem{Hehl:2009es}
F. W. Hehl and B. Mashhoon, 
 ``Formal framework for a nonlocal generalization of Einstein's theory of gravitation",
Phys. Rev. D \textbf{79}, 064028 (2009).  
[arXiv:0902.0560 [gr-qc]]

\bibitem{We}
R. Weitzenb\"ock, 
{\it Invariantentheorie}
(Noordhoff, Groningen, the Netherlands, 1923).

\bibitem{BlHe}
M. Blagojevi\'c and F. W. Hehl, Eds., 
\emph{Gauge Theories of Gravitation} 
(Imperial College Press, London, UK, 2013).

\bibitem{Maluf:2013gaa}
J. W. Maluf, 
``The teleparallel equivalent of general relativity",
Ann. Phys. (Berlin) {\bf 525}, 339-357 (2013).
[arXiv:1303.3897 [gr-qc]]

\bibitem{Aldrovandi:2013wha}
R. Aldrovandi and J. G. Pereira, 
\emph{Teleparallel Gravity: An Introduction}
(Springer, New York, 2013).

\bibitem{Itin:2018dru}
Y. Itin, Y. N.  Obukhov, J. Boos and F. W. Hehl, 
``Premetric teleparallel theory of gravity and its local and linear constitutive law",
Eur. Phys. J. C \textbf{78},  907 (2018).
[arXiv:1808.08048 [gr-qc]]


\bibitem{BeltranJimenez:2019esp}
J.~Beltr{\'a}n Jim{\'e}nez, L.~Heisenberg and T.~S.~Koivisto,
``The Geometrical Trinity of Gravity",
Universe \textbf{5},  no.7, 173 (2019).  \\
doi:10.3390/universe5070173

\bibitem{Capozziello:2022zzh}
S.~Capozziello, V.~De Falco and C.~Ferrara,
``Comparing equivalent gravities: common features and differences",
Eur. Phys. J. C \textbf{82}, no.10, 865 (2022).  \\
doi:10.1140/epjc/s10052-022-10823-x

\bibitem{Cho}
Y. M. Cho, 
``Einstein Lagrangian as the translational Yang-Mills Lagrangian",
Phys. Rev. D {\bf 14}, 2521-2525 (1976).

\bibitem{Hop}
J. Hopkinson, 
``Residual Charge of the Leyden Jar.---Dielectric Properties of different Glasses", 
Phil. Trans. Roy. Soc. London {\bf 167}, 599-626 (1877). 

\bibitem{Poi}
S. Poisson, 
``M\'emoire sur la th\'eorie du magn\'etisme en mouvement",
M\'em. Acad. Sci. France \textbf{6}, 441-570 (1823).

\bibitem{Jack}
J. D. Jackson,
{\it Classical Electrodynamics}, 3rd edn
(Wiley, Somerset, NJ, 1999).

\bibitem{L+L}
L. D. Landau and E. M. Lifshitz, 
\emph{Electrodynamics of Continuous Media} 
(Pergamon,  Oxford, UK, 1960).


\bibitem{HeOb}  
F. W. Hehl and Y. N. Obukhov, 
\emph{Foundations of Classical Electrodynamics: Charge, Flux, and Metric} 
(Birkh\"auser, Boston, MA, USA, 2003). 

\bibitem{VanDenHoogen:2017nyy}
R. J. van den Hoogen, 
``Towards a covariant smoothing procedure for gravitational theories",
J. Math. Phys. \textbf{58}, no.12, 122501 (2017).



\bibitem{Nojiri:2026hij}
S.~Nojiri, S.~Odintsov and V.~K.~Oikonomou,
``Ghost-free non-local F(R) gravity compatible with ACT",
Phys. Lett. B \textbf{874}, 140290 (2026). 
[arXiv:2601.07879 [gr-qc]]  \\
doi:10.1016/j.physletb.2026.140290


\bibitem{Nojiri:2019dio}
S.~Nojiri, S.~D.~Odintsov and V.~K.~Oikonomou,
``Ghost-free non-local $F(R)$ Gravity Cosmology",
Phys. Dark Univ. \textbf{28}, 100541 (2020). 
[arXiv:1911.07329 [gr-qc]]  \\
doi:10.1016/j.dark.2020.100541

\bibitem{Nojiri:2017ncd}
S.~Nojiri, S.~D.~Odintsov and V.~K.~Oikonomou,
``Modified Gravity Theories on a Nutshell: Inflation, Bounce and Late-time Evolution",
Phys. Rept. \textbf{692}, 1-104 (2017).
[arXiv:1705.11098 [gr-qc]]  \\
doi:10.1016/j.physrep.2017.06.001

\bibitem{Deffayet:2025lwl}
C.~Deffayet and R.~P.~Woodard,
``A nonlocal realization of MOND that interpolates from cosmology to gravitationally bound systems",
JCAP \textbf{04},  081 (2026).
[arXiv:2512.10513 [gr-qc]]   \\
doi:10.1088/1475-7516/2026/04/081

\bibitem{Deser:2019lmm}
S.~Deser and R.~P.~Woodard,
``Nonlocal Cosmology II --- Cosmic acceleration without fine tuning or dark energy",
JCAP \textbf{06}, 034 (2019).
[arXiv:1902.08075 [gr-qc]]  \\
doi:10.1088/1475-7516/2019/06/034

\bibitem{Tan:2018bfp}
L.~Tan and R.~P.~Woodard,
``Structure Formation in Nonlocal MOND",
JCAP \textbf{05}, 037 (2018).
[arXiv:1804.01669 [gr-qc]]  \\
doi:10.1088/1475-7516/2018/05/037

\bibitem{Woodard:2018gfj}
R.~P.~Woodard,
``The Case for Nonlocal Modifications of Gravity",
Universe \textbf{4}, no.8, 88 (2018).  [arXiv:1807.01791 [gr-qc]]   \\
doi:10.3390/universe4080088

\bibitem{Maggiore:2014sia}
M. Maggiore and M. Mancarella, 
``Nonlocal gravity and dark energy",
Phys. Rev. D {\bf 90}, no.2, 023005 (2014).  [arXiv:1402.0448 [hep-th]]      \\
\url{https://doi.org/10.1103/PhysRevD.90.023005} 

\bibitem{Balakin:2022gjw}
A.~B.~Balakin and A.~S.~Ilin,
``Self-interaction in a cosmic dark fluid: The four-kernel rheological extension of the equations of state",
Phys. Rev. D \textbf{105}, no.10, 103525 (2022).  [arXiv:2203.16083 [gr-qc]]   \\
doi:10.1103/PhysRevD.105.103525

\bibitem{Koshelev:2022bvg}
A.~S.~Koshelev, K.~S.~Kumar and A.~A.~Starobinsky,
``Non-Gaussianities in generalized non-local R$^{2}$-like inflation",
JHEP \textbf{07}, 094 (2023).   [arXiv:2210.16459 [hep-th]]   \\
doi:10.1007/JHEP07(2023)094

\bibitem{Koshelev:2022olc}
A.~S.~Koshelev, K.~S.~Kumar and A.~A.~Starobinsky,
``Generalized non-local R$^{2}$-like inflation",
JHEP \textbf{07}, 146 (2023).  [arXiv:2209.02515 [hep-th]]    \\
doi:10.1007/JHEP07(2023)146


\bibitem{Jusufi:2023ayv}
K.~Jusufi, A.~Sheykhi and S.~Capozziello,
``Apparent dark matter as a non-local manifestation of emergent gravity",
Phys. Dark Univ. \textbf{42}, 101270 (2023).  [arXiv:2303.14127 [gr-qc]]   \\
doi:10.1016/j.dark.2023.101270

\bibitem{Bajardi:2024kea}
F.~Bajardi and S.~Capozziello,
``Non-Local Cosmology: From Theory to Observations",
Symmetry \textbf{16}, no.5, 579 (2024).  \\
doi:10.3390/sym16050579

\bibitem{Capozziello:2024qol}
S.~Capozziello, A.~Mazumdar and G.~Meluccio,
``Can nonlocal gravity really explain dark energy?",
Phys. Dark Univ. \textbf{45}, 101517 (2024).  [arXiv:2403.11301 [gr-qc]]  \\
doi:10.1016/j.dark.2024.101517

\bibitem{Blome:2010xn}
H.~J.~Blome, C.~Chicone, F.~W.~Hehl and B.~Mashhoon,
``Nonlocal Modification of Newtonian Gravity",
Phys. Rev. D \textbf{81}, 065020 (2010).
[arXiv:1002.1425 [gr-qc]]     \\
doi:10.1103/PhysRevD.81.065020



\bibitem{BMB}
B.~Mashhoon, 
\emph{Nonlocal Gravity}
(Oxford University Press, Oxford, UK, 2017).

\bibitem{Puetzfeld:2019wwo} 
  D. Puetzfeld, Y. N.  Obukhov and F. W. Hehl, 
  ``Constitutive law of nonlocal gravity",
 Phys. Rev. D {\bf 99}, no. 10, 104013 (2019).  [arXiv:1903.04023 [gr-qc]]   \\
 doi:10.1103/PhysRevD.99.104013
  
\bibitem{Mashhoon:2022ynk}
B.~Mashhoon,
``Nonlocal Gravity: Fundamental Tetrads and Constitutive Relations",
Symmetry \textbf{14}, 2116 (2022).
[arXiv:2209.05817 [gr-qc]]

\bibitem{Rahvar:2014yta}
S.~Rahvar and B.~Mashhoon,
``Observational Tests of Nonlocal Gravity: Galaxy Rotation Curves and Clusters of Galaxies",
Phys. Rev. D \textbf{89}, 104011 (2014).  [arXiv:1401.4819 [astro-ph.GA]]   \\
doi:10.1103/PhysRevD.89.104011

\bibitem{Chicone:2015coa}
C.~Chicone and B.~Mashhoon,
``Nonlocal Gravity in the Solar System",
Classical Quantum Gravity \textbf{33}, no.7, 075005 (2016). [arXiv:1508.01508 [gr-qc]]   \\
doi:10.1088/0264-9381/33/7/075005
 
\bibitem{Roshan:2021ljs}
M.~Roshan and B.~Mashhoon,
``Dynamical Friction in Nonlocal Gravity",
Astrophys. J. \textbf{922}, no.1, 9 (2021).  [arXiv:2107.05841 [gr-qc]]  \\
doi:10.3847/1538-4357/ac20d8

          
\bibitem{Roshan:2022zov}
M.~Roshan and B.~Mashhoon,
``Characteristics of Effective Dark Matter in Nonlocal Gravity",
Astrophys. J.  \textbf{934}, no.1, 9  (2022).  [arXiv:2201.12852 [astro-ph.GA]]   \\
doi:10.3847/1538-4357/ac75d6

\bibitem{Roshan:2022ypk}
M.~Roshan and B.~Mashhoon,
``Nonlocal Gravity: Modification of Newtonian Gravitational Force in the Solar System",
Universe \textbf{8}, no.9, 470 (2022).  [arXiv:2205.13276 [gr-qc]]   \\
doi:10.3390/universe8090470

\bibitem{DAgostino:2025wgl}
R.~D'Agostino and V.~De Falco,
``Black hole solutions in the revised Deser-Woodard nonlocal theory of gravity",
JCAP \textbf{07}, 046 (2025). [arXiv:2502.15460 [gr-qc]] \\
\url{https://doi.org/10.1088/1475-7516/2025/07/046}  

\bibitem{DAgostino:2025yej}
R.~D'Agostino and V.~De Falco,
``Quasinormal modes of nonlocal gravity black holes,''
Phys. Rev. D \textbf{112}, no.6, 064028 (2025). [arXiv:2507.01698 [gr-qc]] \\
\url{https://doi.org/10.1103/cyzm-ng1g}  

\bibitem{Borah:2025tvw}
R.~J.~Borah and U.~D.~Goswami,
``Quasinormal modes and shadows of black holes in infinite derivative theory of gravity",
Eur. Phys. J. C \textbf{85}, no.9, 940 (2025). [arXiv:2504.20725 [gr-qc]] \\
\url{https://doi.org/10.1140/epjc/s10052-025-14674-0}   

\bibitem{Tabatabaei:2023qxw}
J.~Tabatabaei, A.~Banihashemi, S.~Baghram and B.~Mashhoon,
``Dynamic dark energy from the local limit of nonlocal gravity",
Int. J. Mod. Phys. D \textbf{32}, no.14, 2342009 (2023).  [arXiv:2305.07630 [gr-qc]]  \\
doi:10.1142/S0218271823420099

\bibitem{Tabatabaei:2023lec}
J.~Tabatabaei, A.~Banihashemi, S.~Baghram and B.~Mashhoon,
``Anisotropic Cosmology in the Local Limit of Nonlocal Gravity",
Universe \textbf{9}, no.9, 377 (2023).  [arXiv:2308.08281 [gr-qc]]  \\
doi:10.3390/universe9090377

\bibitem{Tabatabaei:2022tbq}
J.~Tabatabaei, S.~Baghram and B.~Mashhoon,
``Local limit of non-local gravity: a teleparallel extension of general relativity",
Mon. Not. Roy. Astron. Soc. \textbf{530}, no.1, 795-811 (2024).  [arXiv:2212.05536 [gr-qc]]   \\
doi:10.1093/mnras/stae830

\bibitem{Tabatabaei:2023iwc}
J.~Tabatabaei, A.~Banihashemi, S.~Baghram and B.~Mashhoon,
``Local Limit of Nonlocal Gravity: Cosmological Perturbations",
Astrophys. J. \textbf{965}, no.2, 116 (2024).  [arXiv:2311.07749 [gr-qc]]   \\
doi:10.3847/1538-4357/ad2e04

\bibitem{MaMa}
Z. Mardaninezhad and B. Mashhoon,
``G\"odel's Universe and the Local Limit of Nonlocal Gravity",
Ann. Phys. (Berlin), no.10, e00254 (2025). [arXiv: 2504.14537 [gr-qc]]   \\
\url{https://doi.org/10.1002/andp.202500254} 

\bibitem{Obukhov:2002tm}
Y.~N.~Obukhov and J.~G.~Pereira,
``Metric affine approach to teleparallel gravity",
Phys. Rev. D \textbf{67}, 044016 (2003). [arXiv:gr-qc/0212080 [gr-qc]] \\
\url{https://doi.org/10.1103/PhysRevD.67.044016}  

\bibitem{Obukhov:2019fti}
Y.~N.~Obukhov,
``Exact Solutions in Poincar{\'e} Gauge Gravity Theory",
Universe \textbf{5}, no.5, 127 (2019). [arXiv:1905.11906 [gr-qc]] \\
\url{https://doi.org/10.3390/universe5050127}   

\bibitem{Golovnev:2023uqb}
A.~Golovnev, A.~N.~Semenova and V.~P.~Vandeev,
``Static spherically symmetric solutions in new general relativity",
Classical Quantum Gravity \textbf{41}, no.5, 055009 (2024). [arXiv:2305.03420 [gr-qc]] \\
\url{https://doi.org/10.1088/1361-6382/ad2109}  

\bibitem{Coley:2025cdj}
A.~A.~Coley, N.~T.~Layden and D.~F.~Lopez,
``On black holes in teleparallel torsion theories of gravity",
Gen. Relativ. Gravit. \textbf{57}, no.3, 59 (2025). [arXiv:2503.13215 [gr-qc]] \\
\url{https://doi.org/10.1007/s10714-025-03387-0}   

\bibitem{Lopez:2025gvj}
D.~F.~L{\'o}pez, A.~A.~Coley and R.~J.~van den Hoogen,
``On black holes in new general relativity",
Classical Quantum Gravity \textbf{42}, 24501 (2025).
[arXiv:2508.20314 [gr-qc]]     \\
https://doi.org/10.1088/1361-6382/ae2411

\bibitem{Allahyari:2018cmg}
A.~Allahyari, H.~Firouzjahi and B.~Mashhoon,
``Quasinormal Modes of a Black Hole with Quadrupole Moment",
Phys. Rev. D \textbf{99}, no.4, 044005 (2019). [arXiv:1812.03376 [gr-qc]] \\
\url{https://doi.org/10.1103/PhysRevD.99.044005}  

\bibitem{Allahyari:2019umx}
A.~Allahyari, H.~Firouzjahi and B.~Mashhoon,
``Quasinormal modes of generalized black holes: $\delta$-Kerr spacetime",
Classical Quantum Gravity \textbf{37}, no.5, 055006 (2020).  [arXiv:1908.10813 [gr-qc]]    \\
\url{https://doi.org/10.1088/1361-6382/ab6860}  

\bibitem{Kovetz:2016kpi}
E.~D.~Kovetz, I.~Cholis, P.~C.~Breysse and M.~Kamionkowski,
``Black hole mass function from gravitational wave measurements",
Phys. Rev. D \textbf{95}, no.10, 103010 (2017). [arXiv:1611.01157 [astro-ph.CO]]   \\
\url{https://doi.org/10.1103/PhysRevD.95.103010}  

\bibitem{Tabatabaei:2024juy}
J.~Tabatabaei, S.~Baghram and B.~Mashhoon,
``McVittie{\textendash}Plummer Spacetime: Plummer Sphere Immersed in the FLRW Universe",
Astrophys. J. \textbf{975}, no.2, 240 (2024).  [arXiv:2406.10990 [gr-qc]]   \\
doi:10.3847/1538-4357/ad7b37


\bibitem{Oppenheimer:1939ue}
J.~R.~Oppenheimer and H.~Snyder,
``On Continued gravitational contraction",
Phys. Rev. \textbf{56}, 455-459 (1939).  \\
doi:10.1103/PhysRev.56.455

\bibitem{Datt:1938uwc}
B.~Datt,
``{\"U}ber eine Klasse von L{\"o}sungen der Gravitationsgleichungen der Relativit{\"a}t,''
Z. Phys. \textbf{108}, no.5, 314-321 (1938).  \\
doi:10.1007/BF01374951

\bibitem{Khodabakhshi:2025fmf}
H.~Khodabakhshi, H.~Lu and F.~Shojai,
``Gravitational Collapse: Generalizing Oppenheimer-Snyder and a Conjecture on Horizon Formation Time",
Phys. Rev. D \textbf{112}, no.12, 124057 (2025).
[arXiv:2506.03702 [gr-qc]]  \\
doi:10.1103/dhkq-w5pc


\bibitem{MP}
B. Mashhoon and M. H. Partovi, 
``Gravitational Collapse of a Charged Fluid Sphere", 
Phys. Rev. D \textbf{20}, 2455-2468 (1979). 


\end{thebibliography}
\end{document}